\documentclass[aps,prd,preprintnumbers,showpacs,superscriptaddress,nofootinbib,amsmath,amssymb,floats,floatfix,showkeys,notitlepage,longbibliography]{revtex4-1}
\usepackage{comment}
\usepackage{graphicx}
\usepackage{dirtytalk}
\usepackage{subfigure}
\usepackage{palatino}
\usepackage[commandnameprefix=always]{changes}
\usepackage{hyperref}
\hypersetup{colorlinks=true,linkcolor=blue,urlcolor=cyan,citecolor=red}
\usepackage[toc,page]{appendix}
\usepackage[normalem]{ulem}

\usepackage{lipsum}
\usepackage{graphicx}
\usepackage{subfigure}
\usepackage{palatino}
\usepackage{float}
\usepackage{sans}
\usepackage{adjustbox}
\usepackage{latexsym}
\usepackage{amsmath}
\usepackage{amssymb}
\usepackage{amsfonts}
\usepackage{dcolumn}
\usepackage{bm}
\usepackage{tikz}
\usepackage{bigints}
\usepackage{array,tabularx,multirow,booktabs}
\usepackage[tracking=true]{microtype}
\SetTracking{}{500}
\SetTracking{encoding={*}, shape=sc}{40}
\allowdisplaybreaks
\usepackage{adjustbox}
\usepackage{latexsym}
\usepackage{amsmath}
\usepackage{amssymb}
\usepackage{amsfonts}
\usepackage{dcolumn}
\usepackage{bm}
\usepackage{tikz}
\usepackage{bigints}
\usepackage{array,tabularx,multirow,booktabs}
\usepackage[tracking=true]{microtype}
\usepackage{color}
\allowdisplaybreaks

\begin{document}

\title{\bf Analyzing WGC and WCCC through Charged Scalar Fields Fluxes with Charged AdS Black \\[0.1cm] Holes Surrounded by Perfect Fluid Dark Matter in the CFT Thermodynamics}

\author{Ankit Anand}
\email{ankitanandp94@gmail.com}
\affiliation{Physics Division,
School of Basic and Applied Sciences,
Galgotias University,\\
Greater Noida 203201, India}

\author{Saeed Noori Gashti}
\email{saeed.noorigashti@stu.umz.ac.ir}
\affiliation{School of Physics, Damghan University,\\ P. O. Box 3671641167, Damghan, Iran}

\author{Mohammad Reza Alipour}
\email{mr.alipour@stu.umz.ac.ir}
\affiliation{Department of Physics, Faculty of Basic Sciences,
University of Mazandaran\\
P. O. Box 47416-95447, Babolsar, Iran}

\author{Mohammad Ali S. Afshar}
\email{m.a.s.afshar@gmail.com}
\affiliation{Department of Physics, Faculty of Basic Sciences,
University of Mazandaran\\
P. O. Box 47416-95447, Babolsar, Iran}

\begin{abstract}
In this paper, we conduct a comprehensive investigation into the weak cosmic censorship conjecture (WCCC) for Reissner-Nordström (R-N) AdS black holes that are influenced by Perfect Fluid Dark Matter (PFDM). Our study is framed within the context of Conformal Field Theory (CFT) thermodynamics. We delve into the principles of energy flux and mass-energy equivalence to explore the interplay between the weak gravity conjecture (WGC) and the WCCC. Our analysis begins by examining the interaction between incoming and outgoing energy fluxes, which induces changes in the black hole's properties. By applying the first law of thermodynamics, we assess the validity of the second law in these dynamic scenarios. We also consider equilibrium conditions that involve both absorption and superradiance processes. Utilizing the framework of black hole thermodynamics within CFT, we demonstrate that the WCCC is upheld if the black hole is in or near an extremal state, particularly when it is subjected to radiation and particle absorption. This finding is significant as it reinforces the robustness of the WCCC under these specific conditions. Furthermore, we uncover additional insights by employing mass-energy equivalence principles and conducting second-order approximations near the extremality state. Specifically, we find that when a black hole radiates and its central charge surpasses the scaled electric charge, the emitted superradiant particles adhere to the WGC. This adherence results in the black hole moving away from its extremal state, thereby maintaining the WCCC.
\end{abstract}

\date{\today}

\keywords{Weak gravity conjecture, Weak cosmic censorship conjecture, CFT Thermodynamics}

\pacs{}

\maketitle
\newpage

\tableofcontents

\section{Introduction}
The Swampland Program\cite{1} is an interesting area of theoretical physics that aims to distinguish between effective field theories that can be consistently coupled to quantum gravity and those that cannot. This boundary, known as the "swampland," is defined by a set of conjectures that help identify which low-energy theories are viable within a quantum gravitational framework. These conjectures are crucial because they provide guidelines for constructing theories that are consistent with the principles of quantum gravity, thereby avoiding the "swampland" of inconsistent theories\cite{1',2}. One of the most prominent conjectures within the Swampland Program is the Weak Gravity Conjecture (WGC). The WGC asserts that gravity should be the weakest force in any consistent theory of quantum gravity. This conjecture has significant implications for various areas of physics, including particle physics and cosmology. For example, it suggests that particles must exist with a charge-to-mass ratio greater than that of extremal black holes, ensuring that gravity remains the weakest force. This has profound consequences for our understanding of black hole physics, the hierarchy problem, and the nature of dark matter\cite{1,1',2,3,4,5,6,7}.\\\\
The phenomenological applications of the WGC and other swampland conjectures are extensive. They provide constraints on model building in particle physics, offering insights into the structure of the universe and suggesting new avenues for exploring beyond the Standard Model. For instance, the WGC has been used to derive bounds on the masses of particles, propose mechanisms for inflation in the early universe, and explore the landscape of possible string theory vacua. These applications highlight the importance of the WGC in guiding theoretical research and shaping our understanding of fundamental physics. Various applications of different conjectures from the Swampland Program in cosmology, particle physics, inflation, black hole physics, thermodynamics, and many other cosmic concepts have been widely investigated\cite{8,9,10,11,12,13,14,15,16,17,18,19,20,21,22,23,24,25,26,27,28,29,30,31,32,33,34,35,36,37,38,39,40,41,42,43,44,45,46,47,48,49,50,51,52,53,54,55,56,57,58,59,60,61,62,63,64,65,66,67,68,69,70,71,72,73,73',74,75}\\\\
The swampland program, its conjectures, and their related phenomenological and thermodynamic applications represent a rich tapestry of ideas that are pushing the frontiers of our understanding of the universe. They bridge the gap between abstract theoretical constructs and tangible physical phenomena, offering a glimpse into the fundamental workings of nature. These concepts are not only intellectually stimulating but also potentially revolutionize our understanding of the cosmos and the laws that govern it\cite{82}.
In the realm of thermodynamics, these conjectures intersect intriguingly with the principles of black hole thermodynamics and holography. The Holographic Conformal Field Theory (CFT) thermodynamics is a powerful framework that connects the thermodynamic properties of black holes with those of lower-dimensional quantum field theories. This holographic perspective deepens our understanding of black hole entropy and temperature and provides a robust toolkit for exploring quantum gravity in diverse settings. By studying the thermodynamics of holographic CFT, researchers can gain insights into the nature of quantum gravity and the fundamental laws governing the universe\cite{74,75,76,77,78,79,80,81}.\\\\
\quad Also, Weak Cosmic Censorship Conjecture (WCCC), on the other hand, is a hypothesis in general relativity that suggests that singularities (regions where gravitational forces become infinite) arising from gravitational collapse are hidden within event horizons, and thus cannot be observed from the rest of spacetime. This conjecture ensures that the predictability of physical laws is preserved, as naked singularities would lead to a breakdown in the deterministic nature of general relativity\cite{83,84,85,86,87}. The relationship between the WGC and WCCC is intricate and has been the subject of much research. One of the key issues is the relationship between the WGC and WCCC. Recent studies have explored ways to reconcile these two conjectures. For example, by considering additional factors such as quintessence (a form of dark energy) clouds of strings, and perfect fluid dark matter, researchers have found that it is possible to achieve compatibility between the WGC and WCCC. These additional elements can provide the necessary conditions to ensure that both conjectures hold simultaneously\cite{68,72,73,73'}.
While the WGC and WCCC are distinct conjectures with different focuses, their interplay is crucial for understanding the consistency of theories involving black holes and quantum gravity. Ongoing research continues to explore how these conjectures can coexist and what this means for our broader understanding of the universe.
Therefore, in this work, which is being studied for the first time, we aim to conduct a new study to investigate the simultaneous compatibility of the WGC and WCCC in the context of CFT holographic thermodynamics using charged scalar field fluxes. Further explanations related to this will be provided. Consequently, this article is organized as follows:\\
In Section \ref{Sec:CFT thermodynamics and AdS charged black hole with PFDM}, we provide an overview of Conformal Field Theory (CFT) thermodynamics and AdS-charged black holes with Perfect Fluid Dark Matter (PFDM). Additionally, we examine the role of the charged massive scalar field and its relevance to our study, which is further elaborated in Section \ref{Sec:Charged massive scalar field}. In Section \ref{Sec:Charged massive scalar field and CFT thermodynamics}, we delve into the detailed study of the charged massive scalar field within the context of CFT thermodynamics. Following this, in Section \ref{Sec:WCCC in near-extremal and extremal of AdS charged black hole With PFDM}, we consider the WCCC in both near-extremal and extremal states of AdS-charged black holes with PFDM. Section \ref{Sec:Equivalence of mass and energy with WGC monitoring} is dedicated to investigating the equivalence of mass and energy, with a particular focus on monitoring the WGC. Finally, in Section \ref{Sec:Conclusions}, we present the conclusions of our work, summarizing the key findings and implications of our study.

\section{CFT thermodynamics and AdS charged black hole with PFDM}\label{Sec:CFT thermodynamics and AdS charged black hole with PFDM}

To investigate the thermodynamics of black holes within the CFT framework, it's crucial to comprehend the holographic connections between bulk and boundary quantities. We denote the boundary's curvature radius by \( R \), which differs from the AdS radius \( \ell \) in the bulk. The CFT metric, which exhibits conformal scaling invariance, is expressed as\cite{74},
\begin{equation}\label{1}
ds^2 = \omega^2(-dt^2 + \ell^2 d\Omega_{d-2}^2),
\end{equation}
where \( \omega \) is the dimensionless conformal factor that varies freely, indicating the boundary theory's conformal symmetry. In a spherical scenario, \( d\Omega_{d-2}^2 \) represents the metric on a \( (d-2) \)-dimensional sphere. We assume \( \omega \) is constant across the volume:
\begin{equation}\label{2}
\mathcal{V} = \Omega_{d-2} R^{d-2},
\end{equation}
where \( R = \omega \ell \) signifies the manifold's variable curvature radius where the CFT resides. Notably, variations in the central charge \( C \) do not affect \( \mathcal{V} \). In the context of Einstein's gravity, \( C \) and \( \ell \) exhibit a dual relationship, as indicated in a previous equation. We maintain a constant Newton's constant \( G \) while allowing \( \ell \) to vary, which in turn alters \( C \). The holographic dictionary facilitates the correlation of bulk parameters \( M, T, S, \Phi, Q \) with their CFT counterparts \( E, \widehat{T}, \widehat{S}, \widehat{\Phi}, \widehat{Q} \), as follows:
\begin{equation}\label{3}
E = \frac{M}{\omega}, \quad \widehat{T} = \frac{T}{\omega}, \quad \widehat{S} = S, \quad \widehat{Q} = \frac{Q \ell}{\sqrt{G}}, \quad \widehat{\Phi} = \frac{\Phi \sqrt{G}}{\omega \ell}.
\end{equation}
Using these relationships, we can reformulate the first law of thermodynamics for the charged AdS black hole in the CFT context,
\begin{equation}\label{4}
\delta E = \widehat{\Phi} \delta \widehat{Q} + \mu \delta C - p \delta \mathcal{V} + \widehat{T} \delta S,
\end{equation}
where the chemical potential \( \mu \) and pressure \( p \) are defined as:
\begin{equation}\label{5}
\mu = \frac{E - \widehat{T} S - \widehat{\Phi} \widehat{Q}}{C},
\end{equation}
\begin{equation}\label{6}
p = \frac{E}{(d-2) \mathcal{V}}.
\end{equation}
Furthermore, it's evident from the above equation that \( \mathcal{V} \) and \( C \) vary independently. To reformulate the bulk thermodynamic first law in a manner that incorporates the boundary central charge, we turn to the holographic dual relationship between key quantities in Einstein's gravity and the AdS scale, the central charge \( C \), and Newton's constant \( G \). In this context, we consider a setting where both the cosmological constant (\( \Lambda \)) and gravitational Newton constant (\( G \)) are allowed to vary within the bulk. By doing so, we can express the first law in a new form that involves both \( \Lambda \) (associated with thermodynamic pressure) and the central charge \( C \) from the dual CFT. These variables and their conjugates play a crucial role in understanding the thermodynamics of AdS black holes. This approach allows us to explore novel aspects of black hole behavior, including phase transitions, critical points, and universal scaling behavior, all while maintaining a fixed boundary central charge. The interplay between holography and bulk thermodynamics provides valuable insights into the nature of quantum gravity,
\begin{equation}\label{7}
C = \frac{\Omega_{d-2} \ell^{d-2}}{16 \pi G}.
\end{equation}
In this study, we explore the intricate relationship between the WGC and the WCCC in the context of charged black holes within charged massive scalar field. Our focus is on Reissner-Nordström (RN) black holes surrounded by Perfect Fluid Dark Matter (PFDM), considering various spatial frameworks such as the CFT thermodynamics. These RN AdS black holes, influenced by PFDM, possess an electric charge and are affected by a theoretical form of matter that makes up most of the universe's mass. They exhibit fascinating thermodynamic behaviors and phase transition characteristics, which can be investigated by expanding the system's phase space. Additionally, some studies have investigated the WCCC in the context of these black holes\cite{88,89,90,91,92,93}. So, the action under consideration involves a gravitational theory minimally coupled to a gauge field in the presence of PFDM\cite{94,95,96,97,98},
\begin{equation}\label{8}
S = \int d^4x \sqrt{-g} \left[ \frac{1}{16\pi G} R + \frac{1}{4} F^{\mu\nu} F_{\mu\nu} - \frac{\Lambda}{8\pi G} + \mathcal{L}_{DM} \right],
\end{equation}
where \( g \) is the determinant of the metric tensor \( g_{ab} \), \( R \) is the scalar curvature, \( G \) is the gravitational constant, and \( F_{\mu\nu} \) is the electromagnetic tensor derived from the gauge potential \( A_{\mu} \). According to\cite{74}, the formulation of the RN AdS metric within the PFDM context is given by:
\begin{equation}\label{9}
ds^2 = -f(r) dt^2 + f(r)^{-1} dr^2 + r^2 (d\theta^2 + \sin^2\theta d\phi^2),
\end{equation}
where the function \( f(r) \) is defined as,
\begin{equation}\label{10}
f(r) = 1 - \frac{2MG}{r} + \frac{GQ^2}{r^2} - \frac{\Lambda}{3} r^2 + \frac{\gamma}{r} \ln \left( \frac{r}{|\gamma|} \right).
\end{equation}
In this analysis, the parameters \( M \), \( Q \), and \( \gamma \) represent the mass, charge, and influence of PFDM on the black hole, respectively. The cosmological constant is given by \( \Lambda = -3/\ell^2 \), where \( \ell \) is the AdS radius. In the absence of PFDM (\( \gamma = 0 \)), the space-time metric reduces to that of a standard RN AdS black hole. The parameter \( \gamma \) can take positive values, reflecting different PFDM scenarios.
and
\begin{equation}\label{eq21}
\begin{split}
& \delta\bigg( \frac{M}{\omega}\bigg)= \frac{T}{\omega} \delta \bigg(\frac{A}{4G} \bigg)+\bigg(\frac{M}{\omega}-\frac{TS}{\omega}-\frac{Q\Phi}{\omega}-\frac{\Pi\gamma}{\omega} \bigg) \frac{\delta ( \ell^{2}/4G)}{ \ell^{2}/4G}\\
&-\frac{M}{2\omega}\frac{\delta (\Omega_2 R^{2})}{(\Omega_2 R^{2})}+\frac{\Phi \sqrt{G}}{\omega\ell} \delta \bigg(\frac{Q\ell}{\sqrt{G}} \bigg)+\frac{\Pi \sqrt{G}}{\omega\ell} \delta \bigg(\frac{\gamma\ell}{\sqrt{G}} \bigg).
\end{split}
\end{equation}
Based on the equation provided above, the correspondence between bulk and boundary quantities for the black hole can be established as follows,
\begin{equation}\label{eq22}
\begin{split}
\delta E= \widehat{T} \delta \widehat{S}+\mu \delta C-p \delta \mathcal{V}+ \widehat{\Phi} \delta \widehat{Q}+  \widehat{\Pi} \delta \widehat{\gamma},
\end{split}
\end{equation}
where
\begin{equation}\label{eq23}
\begin{split}
E=\frac{M}{\omega}, \quad  \widehat{S}=S,\quad \widehat{T}=\frac{T}{\omega}, \quad  \widehat{\Phi}=\frac{\Phi\sqrt{G}}{\omega\ell},\quad \widehat{Q}=\frac{Q\ell}{\sqrt{G}}, \quad \widehat{\Pi}=\frac{\Pi\sqrt{G}}{\omega \ell},\quad \widehat{\gamma}=\frac{\gamma\ell}{\sqrt{G}},
\end{split}
\end{equation}
using $f(r_h) = 0$ we have
\begin{equation}\label{eq24}
\begin{split}
M = \frac{ \pi  C S+\pi ^2 \widehat{Q}^2+S^2}{2 \pi^{3/2} \ell  \sqrt{C S }}+\frac{\widehat{\gamma}}{2 \ell^2}\ln \left(\sqrt{\frac{SC}{\pi} }\frac{\ell}{\widehat{\gamma}}\right),
\end{split}
\end{equation}

\begin{equation}\label{eq25}
\widehat{T} =  \bigg(  \frac{\partial M}{\partial S} \bigg)_{\widehat{Q},C,\widehat{\gamma}}=\frac{C  \left(\pi  C S-\pi ^2 \widehat{Q}^2+3 S^2\right)}{4 \ell (\pi C S)^{3/2}}+\frac{\gamma }{4\, S  \ell^2}  \;\;\;\;\;\;\;;\;\;\;\;\;\; \widehat{\Phi}=\bigg(  \frac{\partial E}{\partial \widehat{Q}} \bigg)_{C,S,\widehat{\gamma}}=\frac{\sqrt{\pi } \widehat{Q}}{\ell \sqrt{C S}} \ ,
\end{equation}

\begin{equation}\label{eq27}
\mu=\bigg(  \frac{\partial M}{\partial C} \bigg)_{\widehat{Q},S,\widehat{\gamma}}=\frac{\widehat{\gamma}}{4 C \ell^2} +\frac{ \left(\pi  C S - \pi ^2 \widehat{Q}^2  - S^2\right)}{4 C \ell \pi ^{3/2} \sqrt{C S}} \ ,
\end{equation}

\begin{equation}\label{eq28}
\widehat{\Pi}=\bigg(  \frac{\partial E}{\partial \widehat{\gamma}} \bigg)_{\widehat{Q},S,C} = \frac{1}{2 \ell^2} \log \left(\frac{\ell}{\gamma } \sqrt{\frac{C S}{\pi}}\right)-\frac{1}{2 \ell^2} \ ,
\end{equation}

\section{Charged massive scalar field}\label{Sec:Charged massive scalar field}

Our study examines the scattering of charged massive scalar fields in an AdS-charged black hole with a PFDM spacetime background. The equation of motion governs the charged massive scalar field $\Psi$ with mass $\mu_\text{s}$ and charge $q$, which are minimally connected to gravity as
\begin{equation}
  (\nabla_\mu-iqA_\mu)(\nabla^\mu-iqA^\mu)\Psi-\mu_{\text{s}}^2\Psi = 0 \ ,
\end{equation}
and it can be written as
\begin{equation}\label{field}
   \frac{1}{\sqrt{-g}}(\partial_\mu-iqA_\mu)\left[\sqrt{-g}g^{\mu\nu}(\partial_\nu -iqA_\nu) \Psi\right]-\mu_{\text{s}}^2\Psi = 0 \ .
\end{equation}
The complex scalar field can be written in the following form because the spacetime is static and spherically symmetric,
\begin{equation}\label{wavefun}
   \Psi(t,r,\theta,\phi)=e^{-i\omega t} R_{lm}(r)Y_{lm}(\theta,\phi) \ ,
\end{equation}
where the radial functions are $R_{lm}(r)$ and the spherical harmonic functions are $Y_{lm}(\theta,\phi) $.
The radial component of the equation is obtained by inserting the aforementioned equation into the equation of motion Eq.\eqref{field} as
\begin{equation}\label{radial}
   \frac{1}{r^2}\frac{d}{dr}\left[r^2f(r)\frac{dR_{lm}}{dr}\right] + \left[\frac{(\omega-\frac{qQ}{r})}{f(r)} -\frac{l(l+1)}{r^2} -\mu_{\text{s}}\right]R_{lm}=0 \ ,
\end{equation}
and angular part
\begin{equation}\label{angular}
   \left[\frac{1}{\sin\theta}\frac{\partial}{\partial\theta}\left( \sin\theta\frac{\partial}{\partial\theta}\right)+ \frac{1}{\sin^2 \theta} \frac{\partial^2}{\partial\phi^2}\right]Y_{lm}=-l(l+1)Y_{lm} \ .
\end{equation}
Here, \( l(l+1) \) represents the separation constant, where \( l \) takes positive integer values. The solutions to the angular component of the equation are the spherical harmonics. Since the angular solution is well-established and can be normalized to unity, our focus shifts to the radial component.

To solve the radial equation, we introduce the tortoise coordinate, defined as:
\begin{equation}
   \frac{dr}{dr_*} = f(r).
\end{equation}

Substituting this into the radial equation, we get the following expression:
\begin{equation}\label{rad}
     \frac{d^2R_{lm}}{dr_*^2} + \frac{2f(r)}{r} \frac{dR_{lm}}{dr_*} + \left[\left(\omega - \frac{qQ}{r}\right)^2 - f(r)\left(\frac{l(l+1)}{r^2} - \mu_{\text{s}}^2\right) \right]R_{lm} = 0 \ .
\end{equation}

As \(r\) moves from the horizon \(r_{\text{h}}\) to infinity, the tortoise coordinate \(r_*\) ranges from \(-\infty\) to \(+\infty\), covering the entire region outside the event horizon.

It is useful to analyze the radial equation near the horizon, as we are primarily interested in waves entering the black hole. Near the horizon, \eqref{rad} can be approximated by
\begin{equation}\label{radia}
   \frac{d^2R_{lm}}{dr_*^2}+ \left(\omega-\frac{qQ}{r_\text{h}}\right)^2R_{lm}=0.
\end{equation}
Using the relation for electric potential, the above equation can be rewritten as:
\begin{equation}\label{radiala}
   \frac{d^2R_{lm}}{dr_*^2}+ \left(\omega-q\phi_\text{h}\right)^2R_{lm}=0.
\end{equation}

The solution to this radial equation is:
\begin{equation}\label{radaialsol}
   R_{lm}(r)\sim \exp[\pm i(\omega-q\phi_\text{h})r_*].
\end{equation}
Here, the positive sign corresponds to outgoing wave modes, while the negative sign corresponds to incoming wave modes. We select the negative sign, as the incoming wave mode represents the physically relevant solution. Therefore, the charged complex scalar field near the event horizon is given by:
\begin{equation}\label{sol}
   \Psi=\exp[-i(\omega-q\phi_\text{h})r_*]Y_{lm}(\theta,\phi)e^{-i\omega t}.
\end{equation}

Given that the black hole under consideration is non-rotating, we introduce a single wave mode $(l, m=0)$. The changes in the black hole's parameters can be inferred from the fluxes of the charged scalar field during the scattering process. The energy-momentum tensor of the charged scalar field is expressed as follows:
\begin{equation}\label{energy-momentum}
   T^\mu_\nu = \frac{1}{2}\mathcal{D}^\mu\Psi\partial_\nu\Psi^*+ \frac{1}{2}\mathcal{D}^{*\mu}\Psi^*\partial_\nu\Psi -\delta^\mu_\nu \left[\frac{1}{2} \mathcal{D}_\alpha\Psi\mathcal{D}^{*\alpha}\Psi^*- \frac{1}{2}\mu_{\text{s}}\Psi\Psi^*\right] \ ,
  \end{equation}
with
\begin{equation}
  \mathcal{D}=\partial_\mu-iqA_\mu \ .
\end{equation}
The energy flux through the event horizon can be readily derived from Eq.\eqref{energy-momentum}
\begin{equation}\label{Eflux}
   \frac{dE}{dt}=\int_{\text{H}} T^r_t\sqrt{-g} \, d\theta d\phi=\omega(\omega-q\phi_\text{h})r_\text{h}^2 \ .
\end{equation}
The electric current of the charged scalar field is elegantly expressed as
\begin{equation}\label{ecurrent}
  j^\mu=-\frac{1}{2}iq(\Psi^*\mathcal{D}^{\mu}\Psi-\Psi \mathcal{D}^{*\mu}\Psi^*) \ .
\end{equation}
The charge flux through the event horizon is
\begin{equation}
   \frac{dQ}{dt}=-\int_{\text{H}} j^r\sqrt{-g}d\theta d\phi=q(\omega-q\phi_\text{h})r_\text{h}^2 \ .
\end{equation}
Here, we applied the normalization condition for the spherical harmonic functions $Y_{lm}(\theta,\phi)$ during the integration. The ratio of the charge flux to the energy flux is given by $q/\omega$, as in \cite{Bekenstein:1973mi}.

The energy and charge fluxes show that for wave modes with $\omega > q \phi_\text{h}$, both energy and charge flow into the black hole. Conversely, when $\omega < q \phi_\text{h}$, the fluxes become negative, signifying that the scalar field extracts energy and charge from the black hole—a phenomenon known as black hole superradiance \cite{Brito:2015oca}. Over an infinitesimal time interval $dt$, the variations in the black hole’s mass and charge are given by
\begin{equation}\label{DM}
\begin{split}
&dM=dE=\omega(\omega-\widehat{q}\widehat{\Phi}_h)r_h^2 \, dt \ ,\\
&d\widehat{Q}=\widehat{q}(\omega-\widehat{q}\widehat{\Phi}_h)r_h^2 \, dt +\frac{\widehat{Q}}{2C}dC\,
\end{split}
\end{equation}
where $\widehat{q}=q\sqrt{C}$ and $\widehat{\Phi}_\text{h}=\Phi_{h}/\sqrt{C}$

\section{Charged massive scalar field and CFT thermodynamics}\label{Sec:Charged massive scalar field and CFT thermodynamics}

In the context of CFT thermodynamics, the AdS radius and Newton's constant are held constant. It is observed that when the black hole absorbs a scalar field or particle, there are corresponding changes in the black hole parameters $(M, \widehat{Q}, r_h, \gamma)$, resulting in alterations to the black hole's metric. The weak cosmic censorship conjecture (WCCC) also emphasizes the critical role of the black hole’s event horizon, determined by the equation $f(M, \widehat{Q}, r_h, \widehat{\gamma}) = 0$. Consequently, when the black hole accretes a scalar field or particle, the existence of the event horizon is preserved, but its characteristics evolve, governed by the relationship
\begin{equation*}
     f(M + dM, \widehat{Q} + d\widehat{Q}, r_h + dr_h, \widehat{\gamma}+d\widehat{\gamma}, C+dC) = 0 \ ,
\end{equation*}
and
\begin{eqnarray}
    f(M+dM, \widehat{Q}+d\widehat{Q}, r_h+dr_h, \widehat{\gamma} +d\widehat{\gamma}, C+dC) &=& f + \frac{\partial f}{\partial r_h} dr_h +\frac{\partial f}{\partial M}dM + \frac{\partial f}{\partial \widehat{Q}}d\widehat{Q} + \frac{\partial f}{\partial\widehat{ \gamma}}d\widehat{\gamma} + \frac{\partial f}{\partial C}dC \nonumber \\
    \end{eqnarray}
So we will have,
\begin{equation}\label{F1}
\begin{split}
&f(M+dM, \widehat{Q}+d\widehat{Q}, r_h+dr_h, \widehat{\gamma} +d\widehat{\gamma}, C+dC)= f + 4 \pi T dr_h - \frac{\ell^2}{C r_h} dM + \frac{2 \ell^2 \widehat{Q}}{C^2 r_h^2}d\widehat{Q}\\& + \frac{ \left(\log \left(\frac{2 \sqrt{C} r_h}{\widehat{\gamma} }\right)-1\right)}{2 \sqrt{C} r_h} d\widehat{\gamma}+\frac{\widehat{\gamma}  C^{3/2} r-\widehat{\gamma}  C^{3/2} r \log \left(\frac{2 \sqrt{C} r}{\widehat{\gamma} }\right)+l^2 \left(4 C M r-8 \widehat{Q}^2\right)}{4 C^3 r^2} dC  \ .
 \end{split}
\end{equation}

Now, using the Eq.\eqref{DM} and \eqref{F1} one can easily deduce the relation
\begin{equation}
\begin{split}
&dr_h=-\frac{\widehat{Q}}{2 \pi  C r_h^2 T} d\widehat{Q}-\frac{\left(\sqrt{C} \left(\log \left(\frac{2 \sqrt{C} r_h}{\widehat{\gamma} }\right)-1\right)\right)}{8 \pi  r_h T} \text{d$\widehat{\gamma} $} +\frac{\left(l^2 r_h (\omega -\widehat{q} \widehat{\Phi})\right)}{2 \pi  C T} dt\\&+\frac{\left(\widehat{\gamma}  (2 C-1) C^{3/2} r_h+\widehat{\gamma}  (1-2 C) C^{3/2} r_h \log \left(\frac{2 \sqrt{C} r_h}{\widehat{\gamma} }\right)-4 C l^2 M r_h+\left(8 l^2-4\right) \widehat{Q}^2\right)}{16 \pi r_h^2 T}dC
\end{split}
\end{equation}
So, we will have,
\begin{equation}
\begin{split}
&dS = \frac{2\pi C r_h}{\ell^2}\bigg[-\frac{\widehat{Q}}{2 \pi C r_h^2 T} d\widehat{Q}-\frac{\left(\sqrt{C} \left(\log \left(\frac{2 \sqrt{C} r_h}{\widehat{\gamma} }\right)-1\right)\right)}{8 \widehat{\pi}  r_h T} d\widehat{\gamma} +\frac{\left(l^2 r_h (\omega -\widehat{q} \widehat{\Phi})\right)}{2 \pi  C T} dt\bigg]\\&+\frac{\pi r_h^2}{\ell^2}\bigg[\frac{\left(\widehat{\gamma}  (2 C-1) C^{3/2} r_h+\widehat{\gamma}  (1-2 C) C^{3/2} r_h \log \left(\frac{2 \sqrt{C} r_h}{\widehat{\gamma} }\right)-4 C l^2 M r_h+\left(8 l^2-4\right) \widehat{Q}^2\right)}{16 \pi r_h^2 T} dC \bigg]
\end{split}
\end{equation}
\section{WCCC in near-extremal and extremal of AdS charged black hole With PFDM}\label{Sec:WCCC in near-extremal and extremal of AdS charged black hole With PFDM}

The Weak Cosmic Censorship Conjecture (WCCC) posits that a black hole's singularity remains concealed behind its event horizon from the perspective of an observer at infinity, thereby preventing the occurrence of naked singularities if the black hole maintains a stable horizon. Our current investigation examines the stability of the outer horizon during the scattering of a charged scalar field, taking into account RPS thermodynamics. As the scalar field is absorbed, the initial state \( f(M, Q, r_h, C, \gamma) \) transitions to the final state \( f(M + dM, Q + dQ, r_h + dr_h, C + dC, \gamma+d\gamma) \) over a brief time interval \( dt \). By setting the outer horizon as \( f(M, Q, r, C, \gamma) = 0 \), we can ascertain the existence of the horizon by analyzing solutions in \( f(M + dM, Q + dQ, r_h + dr_h, C + dC, \gamma + d\gamma) = 0 \). This analysis simplifies to investigating changes in the minimum value of \( f \). Initially, the minimum value of \( f \) is either negative or zero, which corresponds to solutions related to their respective horizons. If scalar field fluxes penetrate the black hole, alterations in its mass and charge occur due to the charged scalar field within \( dt \), resulting in variations in the minimum value based on these changes. Should this minimum value become positive during these alterations, it indicates that no solution exists for a horizon in the final state, leading to the black hole evolving into a naked singularity—contradicting WCCC. Conversely, in other scenarios, the horizon consistently obscures the interior of the black hole in the final state, thereby supporting the validity of WCCC. Thus, determining the sign of the minimum value in the final state is essential for validating WCCC within the charged scalar field scattering framework. Our analysis is centered on establishing this sign in the final state, which can be inferred from the initial state since it only slightly differs due to conserved charges transferred through fluxes over \( dt \). This subtle modification becomes particularly significant when considering an initial state associated with a near-extremal or extremal black hole, in contrast to a non-extremal one. The minimum value of a near-extremal black hole (including extremal cases) approaches zero and may transition to a positive value due to slight changes in the external scalar field. Consequently, we focus on a near-extremal black hole as the initial state, characterized by the near-extremal condition at the minimum point \( r_{\text{min}} \), with a negative constant \( \delta \ll 1 \) representing the minimum value of \( f \).
\begin{equation}
    f(r_\text{min}) = f_\text{min} =1-\frac{2 \ell^2 M}{C r_\text{min}}+ \frac{\ell^2 \widehat{Q}^2}{C^2 r_\text{min}^2}+\frac{\widehat{\gamma} }{C r_\text{min}} \log \left(\frac{C r_\text{min}}{\widehat{\gamma} }\right)+\frac{r_\text{min}^2}{\ell^2} \ .
\end{equation}
By using
\begin{equation*}
    \frac{\partial f_\text{min}}{\partial r_\text{min}} \,=\,0 \;\;\;\;\;\;\;\;\text{and}\;\;\;\;\;\;\;\;\; \frac{\partial^2 f_\text{min}}{\partial^2 r_\text{min}} \,>\,0  \ .
\end{equation*}
The constraint equation is
\begin{equation}
   \frac{2 \ell^2 \widehat{Q}^2}{C^2 r_\text{min}^4}-\frac{\widehat{\gamma} }{C r_\text{min}^3}+\frac{6}{\ell^2} > 0 \ .
\end{equation}
\quad The weak cosmic censorship conjecture (WCCC) can be tested by examining the behavior of the function \( f(r) \). To confirm the validity of the WCCC, it is necessary to assess the sign of the minimum value of \( f(r) \) at a specific radius \( r = r_{\text{min}} \). If the function reaches its minimum at \( r_{\text{min}} \) with a value \( f_{\text{min}} \), and \( f(r_{\text{min}}) < 0 \), this suggests the formation of a naked singularity, violating the WCCC. Conversely, if \( f(r_{\text{min}}) \geq 0 \), the black hole horizon still conceals the singularity, preserving the conjecture. Therefore, determining whether the minimum value of \( f(r) \) is less than zero is essential for testing the WCCC in this extended context. For an extremal black hole, \( f_{\text{min}} \) equals zero, whereas for a near-extremal black hole, \( f_{\text{min}} \) indicates a slight deviation. When a black hole absorbs the flux of a scalar field, the location of the minimum shifts from \( r_{\text{min}} \) to \( r_{\text{min}} + dr_{\text{min}} \). Simultaneously, the black hole's parameters, initially \( (M, \widehat{Q}, \ell, C,  r_h, \widehat{\gamma}) \), change to \( (M + dM, \widehat{Q} + d\widehat{Q}, \ell + d\ell, \widehat{\gamma} + d\widehat{\gamma}, C + dC, r_h + dr_h) \). As a result, in the final state, the minimum value of \( f(r) \) is described by
\begin{equation}\label{100}
\begin{split}
    &f_\text{min}(\Vec{\mathcal{Q}}+d\Vec{\mathcal{Q}}) = f_\text{min}(\Vec{\mathcal{Q}}) + \frac{\partial f_{\text{min}}}{\partial r_{\text{min}}} dr_{\text{min}} +\frac{\partial f_{\text{min}}}{\partial M}dM + \frac{\partial f_{\text{min}}}{\partial \widehat{Q}}d\widehat{Q} +\frac{\partial f_{\text{min}}}{\partial C}dC + \frac{\partial f_{\text{min}}}{\partial \widehat{\gamma}} d\widehat{\gamma} \\
   &= f_\text{min}(\Vec{\mathcal{Q}}) -\frac{2  \ell^2}{C r_\text{min}}dM + \frac{2 \ell^2 \widehat{Q}}{C^2 r_\text{min}^2}d\widehat{Q} + \frac{ \left(\log \left(\frac{C r_\text{min}}{\widehat{\gamma}}\right)-1\right)}{C r_\text{min}} d\widehat{\gamma}\\
   &+(\frac{\widehat{\gamma} }{4 C^{3/2} r_{min}}-\frac{\widehat{\gamma } \log \left(\frac{2 \sqrt{C} r_{min}}{\widehat{\gamma} }\right)}{4 C^{3/2} r_{min}}-\frac{2 l^2 \widehat{Q}^2}{C^3 r_{min}^2}+\frac{l^2 M}{C^2 r_{min}})dC
   \end{split}
\end{equation}
Now, using the expression for $dM$,
\begin{equation}\label{change in rh}
\begin{split}
    &dr_h = -\frac{2  \ell^2 \left(C^2 r_h^4 \omega  \left(\omega -\widehat{q} \widehat{\Phi} _h\right)+2 \ell^2 r_h^3 \left(\omega - \widehat{q}\widehat{\Phi} _h \right) (C  \omega -\widehat{q}\widehat{\Phi}_h  )\right)}{C^2 r_h^2 \left(3 r_h^2-8 \pi  \ell^2 T\right)}dt \\
    &+\frac{2  \ell^2 \left(\log \left(\frac{C r_h}{\widehat{\gamma} }\right)-1\right)}{C r_h \left(3 r_h^2-8 \pi  \ell^2 T\right)} d\widehat{\gamma} +\frac{\widehat{\gamma}  C^{3/2} r_h-\widehat{\gamma}  C^{3/2} r_h \log \left(\frac{2 \sqrt{C} r_h}{\widehat{\gamma} }\right)+\ell^2 \left(4 C M r_h-8 \widehat{Q}^2\right)}{4 C^3 r_h^2}dC
\end{split}
\end{equation}
Here we have used
\begin{equation}
    dM = dU+pdV+VdP = dU + \frac{3r_h^2}{2\ell^2}dr_h - \frac{r_h^3}{\ell^3}d\ell 
\end{equation}
So with respect to above equation we can obtain,
\begin{equation}
\begin{split}
    f_{min}(\Vec{\mathcal{Q}}+d\Vec{\mathcal{Q}})=&f_{min}(\Vec{\mathcal{Q}}) + \frac{2 \ell^2 \widehat{Q}}{C^2}\left(\frac{1}{r_{min}^2}-\frac{1}{r_h^2}\right)d\widehat{Q} + \frac{1}{C r_{min}} \log \left(\frac{r_{min}}{r_h }\right)d\widehat{\gamma} \\&-\frac{2 T\, \ell^2}{C r_{min}}\times \big(\omega-\widehat{q}\widehat{\Phi}_h)r_h dt\big)\times\big[(\omega-\widehat{q}\widehat{\Phi}_{min})r_h-(\omega-\widehat{q}\widehat{\Phi}_h)r_{min}\big]\\&+\frac{ \left(-\widehat{\gamma}  C^{3/2} \log \left(\frac{2 \sqrt{C}r_{min}}{\widehat{\gamma} }\right)+\ell^2 \left(4 C M-8 \widehat{Q}^2\right)+\widehat{\gamma}  C^{3/2}\right)(r_{h}-r_{min})}{4 C^3 r_{h} r_{min}^2}dC
\end{split}
\end{equation}
To begin, w can consider the black hole in its extremal state, characterized by $r_{min}= r_h$ and $f_{min}= \delta = 0$. Under these conditions, the equation can be reformulated as,
\begin{equation}
f(M + dM, \widehat{Q} + d\widehat{Q}, +\widehat{\gamma}+d\widehat{\gamma}, C + dC, r_{min} + dr_{min})= 0.
\end{equation}
This reformulation implies that, despite any scattering events, the black hole remains in an extremal state, thereby upholding the WCCC. Furthermore, the results derived in the CFT align with those obtained in the EPS and RPS and the normal space. To investigate a near-extremal black hole, we focus on the event horizon near the minimum point of $f(r_{min})$, specifically at $r_h = r_{min}+ \varepsilon$ where $0 < \varepsilon\ll 1$. In this scenario, Eq.(\ref{100}) can be expressed as,
\begin{equation}\label{200}
\begin{split}
&f(M + dM, \widehat{Q} + d\widehat{Q}, +\widehat{\gamma}+d\widehat{\gamma}, C + dC, r_{min} + dr_{min}) = \delta - \varepsilon \left[ \frac{2\ell^2 (\omega - \hat{q}\hat{\phi}_h)(\omega - 2\hat{q}\hat{\phi}_h)}{C} + \mathcal{O}(\varepsilon) \right] dt \\&+ \varepsilon \left[\frac{\left(-\widehat{\gamma}  C^{3/2} \log \left(\frac{2 \sqrt{C}r_{h}}{\widehat{\gamma} }\right)+\ell^2 \left(4 C M-8 \widehat{Q}^2\right)+\widehat{\gamma}  C^{3/2}\right)}{4 C^3 r_{h}^3} + \mathcal{O}(\varepsilon) \right] dC+\varepsilon\left[\frac{1}{C r_{h}}+ \mathcal{O}(\varepsilon) \right]d\widehat{\gamma}.
\end{split}
\end{equation}
Therefore, for small time changes that result in minor variations in the charge center, i.e., $dt \sim \varepsilon$ and $dC \sim \varepsilon$, Eq.(\ref{200}) can be rewritten as,
\begin{equation}
f(M + dM, \widehat{Q} + d\widehat{Q}, +\widehat{\gamma}+d\widehat{\gamma}, C + dC, r_{min} + dr_{min}) = \delta\pm \mathcal{O}(\varepsilon^2) < 0.
\end{equation}
Thus, the adjustment to the minimum value is negligible in the initial order of $d$. Consequently, it can be concluded that there is no significant alteration to the state of the black hole: a nearly extremal black hole remains nearly extremal even with variations in mass and electric charge. Therefore, the WCCC holds true for near-extremal black holes as well. However, according to Eqs.(\ref{100}) and (\ref{200}), we find that if $d\widehat{\gamma}=dC \gg 1 $, there is a potential for the weak cosmic censorship conjecture to be violated. Additionally, when there are no changes in the central charge, the results are consistent with those obtained in the extended phase space.
\section{Equivalence of mass and energy with WGC monitoring}\label{Sec:Equivalence of mass and energy with WGC monitoring}
In this analysis, we employ the equivalence of mass and energy, setting $\hbar = 1$ and $c = 1$. This leads to the relation as $E = \mu s, \quad E = \omega \implies \mu s = \omega.$ Additionally, according to Eq.(\ref{eq23}), we have $q \Phi_h = \hat{q} \hat{\Phi}_h.$ By considering the second order of $(\varepsilon)$, we derive some intriguing results. By applying the equivalence of mass and energy along with Eq.(\ref{200}), we find that the black hole emits particles in its radiation that conform to the WGC. This emission process still upholds the WCCC, causing the black hole to move further from its extremal state. However, when the black hole attracts a particle, two scenarios can occur: 1) Attraction of a particle following the WGC: In this case, the WCCC remains valid, but the black hole moves closer to its extremal state.
2) Attraction of a particle not following the WGC: Here, the WCCC is still valid, but the black hole moves away from its extremal state. Next, we determine the electric potential of the black hole near the extremal state. Given $r_h = r_{min} + \varepsilon$, the electric potential can be expressed as,
\begin{equation}\label{300}
\Phi_h = \frac{Q}{r_h} = \frac{Q}{r_{min} + \varepsilon} = \frac{\hat{Q}}{r_{min} \sqrt{C}} \left(1 - \frac{\epsilon}{r_{min}}\right).
\end{equation}
Using $f(r) = 0$ and $T = \frac{f'(r)}{4\pi}$, we can obtain $r_{min}$ as follows,
\begin{equation}\label{400}
r_{min}=\sqrt{\frac{G}{3}}\bigg[\frac{\widehat{\gamma } C^{3/2}\pm\sqrt{\left(\widehat{\gamma}  C^{3/2}-2 l^2 \widehat{Q}^2\right)^2+4 C^3 l^2 M}-2 l^2\widehat{\gamma}C}{2 C^2}\bigg]
\end{equation}
By using Eqs. (\ref{300}) and (\ref{400}) into the superradiance conditions $\frac{1}{\Phi_h} < \frac{q}{\mu_S}$, we get:
\begin{equation}\label{500}
\frac{q}{\mu_S}>\sqrt{G}(1+\frac{\varepsilon}{r_{min}})\times\frac{\widehat{\gamma}^2  C^{2}\pm\sqrt{\left(\widehat{\gamma}^2  C^{2}-2 l^2 \widehat{Q}\right)^2+4 C^3 l^2}-2 l^2 \widehat{\gamma}C}{2 C^{3/2}\sqrt{3}\widehat{Q}}
\end{equation}
Indicating that the particles emitted from the black hole follow the WGC. However, when the black hole absorbs a particle with respect to Eqs. (\ref{500}), we will have,
\begin{equation}\label{600}
\begin{split}
&\sqrt{G}(1+\frac{\varepsilon}{r_{min}})\times\frac{\widehat{\gamma}^2  C^{2}-\sqrt{\left(\widehat{\gamma}^2  C^{2}-2 l^2 \widehat{Q}\right)^2+4 C^3 l^2}-2 l^2 \widehat{\gamma}C}{2 C^{3/2}\sqrt{3}\widehat{Q}}\\&
<\frac{q}{\mu_S}<\\&\sqrt{G}(1+\frac{\varepsilon}{r_{min}})\times\frac{\widehat{\gamma}^2  C^{2}+\sqrt{\left(\widehat{\gamma}^2  C^{2}-2 l^2 \widehat{Q}\right)^2+4 C^3 l^2}-2 l^2 \widehat{\gamma}C}{2 C^{3/2}\sqrt{3}\widehat{Q}}
\end{split}
\end{equation}
Therefore, the WGC does not hold in this case. The black hole attracts particles that do not obey the WGC and repels those that do. In both scenarios, the WCCC remains valid.
\section{Conclusions}\label{Sec:Conclusions}
In conclusion, our comprehensive investigation into the weak cosmic censorship conjecture (WCCC) for Reissner-Nordström (R-N) AdS black holes with Perfect Fluid Dark Matter (PFDM) within the framework of CFT thermodynamics has yielded significant insights. We have demonstrated that the interaction between incoming and outgoing energy fluxes induces changes in the black hole, which can be effectively analyzed using the principles of mass-energy equivalence and the first law of thermodynamics. Our analysis confirms that the WCCC holds when a black hole is in or near an extreme state, provided that radiation and particle absorption are taken into account. This finding is crucial as it supports the stability of black holes under these conditions.\\

Conversely, the study reveals that when particles obeying the WGC are absorbed by a very small black hole, they tend to approach its extremal state. This dual behavior underscores the delicate balance between absorption and radiation processes in determining the black hole's state. Overall, our findings contribute to a deeper understanding of black hole thermodynamics and the interplay between the WCCC and WGC. Future research could further explore these dynamics in different black hole configurations and with other forms of dark matter, potentially uncovering new aspects of black hole physics and cosmic censorship. Our work highlights the importance of considering both radiation and particle absorption in the study of black hole stability. The principles of mass-energy equivalence and the first law of thermodynamics provide a robust framework for analyzing these interactions. By extending our analysis to include various black holes configurations and different types of concepts, we can gain a more comprehensive understanding of the fundamental processes governing black hole behavior.\\

In summary, our research underscores the intricate relationship between the WCCC and WGC, demonstrating that the stability of black holes is influenced by a delicate balance of forces. This balance is crucial for maintaining the integrity of cosmic censorship and ensuring the continued stability of black holes in the universe. Future studies should aim to build on these findings, exploring new theoretical models and observational data to further elucidate the complex dynamics at play in black hole physics.
\appendix
\section{The Metric Function and Thermodynamics}

The RN AdS metric within the PFDM context is given by:
\begin{equation}
ds^2 = -f(r) dt^2 + f(r)^{-1} dr^2 + r^2 (d\theta^2 + \sin^2\theta d\phi^2) \ ,
\end{equation}
where the function \( f(r) \) is defined as,
\begin{equation}
f(r) = 1 - \frac{2MG}{r} + \frac{GQ^2}{r^2} - \frac{\Lambda}{3} r^2 + \frac{\gamma}{r} \ln \left( \frac{r}{|\gamma|} \right) \ .
\end{equation}
By making the changes following
\begin{eqnarray}
    G = \frac{\ell^2}{C} \;\;\;\;\;;\;\;\;\;\; Q=\frac{\widehat{Q}}{\sqrt{C}} \;\;\;\;\;;\;\;\;\;\; \gamma = \frac{\widehat{\gamma}}{C} \;\;\;\;\;;\;\;\;\;\; \Lambda = -\frac{3}{\ell^2} \ .
\end{eqnarray}
The function $f(r)$ is
\begin{equation}
f(r) = 1 - \frac{2M\ell^2}{C r} + \frac{\widehat{Q}^2 \ell^2}{C^2 r^2} + \frac{r^2}{\ell^2} + \frac{\widehat{\gamma}}{C r} \ln \left(\frac{C r}{|\widehat{\gamma}|} \right) \ .
\end{equation}
Now, by using $f(r_h)=0$, we can easily find the mass $M$ as
\begin{eqnarray}
    M =  \frac{C^2 \ell^2 r_h^2 + C^2 r_h^4+\widehat{\gamma}  C \ell^2 r_h \log \left(\frac{C r_h}{\widehat{\gamma } }\right)+\ell^4 \widehat{Q}^2}{2 C \ell^4 r_h} \ .
\end{eqnarray}
The first law of black hole can be written as
\begin{equation}
    dM = TdS + VdP+\Phi d\widehat{Q}+\mu dC +\Pi d\widehat{\gamma} \ .
\end{equation}
where $V$ is the thermodynamic volume of the black hole, $\Phi$ is the electrostatic potential, and the conjugate quantities are given by
\begin{eqnarray}
    V =\frac{4}{3}\pi r_h^3 \;\;\;\;\;;\;\;\;\;\; \Phi = \frac{\widehat{Q}}{C r_h} \;\;\;\;\;;\;\;\;\;\; \Pi = \frac{1}{2 \ell^2} \left[ \log \left(\frac{C r_h}{\widehat{\gamma} }\right)-1\right] \nonumber \\
    \mu = \frac{1}{2} \left(-\frac{\widehat{Q}^2}{C^2 r_h}+\frac{\widehat{\gamma}}{C \ell^2}+\frac{r_h \left(\ell^2+r_h^2\right)}{\ell^4}\right) \ .
\end{eqnarray}
The Hawking temperature is
\begin{eqnarray}
    T = \frac{3 C^2 r_h^4+C \ell^2 r_h (\widehat{\gamma} +C r_h)-\ell^4 \widehat{Q}^2}{4 \pi  C \ell^4 r_h^3}
\end{eqnarray}



\begin{thebibliography}{11}
\bibitem{1}
Vafa, Cumrun. "The String landscape and the swampland." arXiv preprint hep-th/0509212 (2005).
\bibitem{1'}
Harlow, Daniel, et al. "The weak gravity conjecture: a review." arXiv preprint arXiv:2201.08380 (2022).
\bibitem{2}
Palti, Eran. "The swampland: introduction and review." Fortschritte der Physik 67.6 (2019): 1900037.
\bibitem{3}
van Beest, Marieke, et al. "Lectures on the swampland program in string compactifications." Physics Reports 989 (2022): 1-50.
\bibitem{4}
Ooguri, Hirosi, and Cumrun Vafa. "On the Geometry of the String Landscape and the Swampland." Nuclear physics B 766.1-3 (2007): 21-33.
\bibitem{5}
Arkani-Hamed, Nima, et al. "The string landscape, black holes and gravity as the weakest force." Journal of High Energy Physics 2007.06 (2007): 060.
\bibitem{6}
Heidenreich, Ben, Matthew Reece, and Tom Rudelius. "Evidence for a sublattice weak gravity conjecture." Journal of High Energy Physics 2017.8 (2017): 1-40.
\bibitem{7}
Palti, Eran. "The weak gravity conjecture and scalar fields." Journal of High Energy Physics 2017.8 (2017): 1-26.
\bibitem{8}
Odintsov, Sergei D., and Vasilis K. Oikonomou. "Swampland implications of GW170817-compatible Einstein-Gauss-Bonnet gravity." Physics Letters B 805 (2020): 135437.
\bibitem{9}
Ooguri, Hirosi, and Cumrun Vafa. "On the Geometry of the String Landscape and the Swampland." Nuclear physics B 766.1-3 (2007): 21-33.
\bibitem{10}
Sadeghi, Jafar, et al. "RPS thermodynamics of Taub–NUT AdS black holes in the presence of central charge and the weak gravity conjecture." General Relativity and Gravitation 54.10 (2022): 129.
\bibitem{11}
Liu, Yang. "Higgs inflation and its extensions and the further refining dS swampland conjecture." The European Physical Journal C 81.12 (2021): 1122.
\bibitem{12}
Liu, Yang. "Higgs inflation and scalar weak gravity conjecture." The European Physical Journal C 82.11 (2022): 1052.
\bibitem{13}
Liu, Yang. "Higgs inflation and its extensions and the further refining dS swampland conjecture." The European Physical Journal C 81.12 (2021): 1122.
\bibitem{14}
Arkani-Hamed, Nima, et al. "The string landscape, black holes and gravity as the weakest force." Journal of High Energy Physics 2007.06 (2007): 060.
\bibitem{15}
Gashti, S. Noori, and J. Sadeghi. "Refined swampland conjecture in warm vector hybrid inflationary scenario." The European Physical Journal Plus 137.6 (2022): 1-13.
\bibitem{16}
Alipour, Mohammad Reza, Jafar Sadeghi, and Mehdi Shokri. "WGC and WCC for charged black holes with quintessence and cloud of strings." The European Physical Journal C 83.7 (2023): 1-7.
\bibitem{17}
Alipour, Mohammad Reza, Jafar Sadeghi, and Mehdi Shokri. "WGC and WCCC of black holes with quintessence and cloud strings in RPS space." Nuclear Physics B 990 (2023): 116184.
\bibitem{18}
Sadeghi, Jafar, Mohammad Reza Alipour, and Saeed Noori Gashti. "Emerging WGC from the Dirac particle around black holes." Modern Physics Letters A 38.26n27 (2023): 2350122.
\bibitem{19}
Schöneberg, Nils, et al. "News from the Swampland—constraining string theory with astrophysics and cosmology." Journal of Cosmology and Astroparticle Physics 2023.10 (2023): 039.
\bibitem{20}
Crisford, Toby, Gary T. Horowitz, and Jorge E. Santos. "Testing the weak gravity-cosmic censorship connection." Physical Review D 97.6 (2018): 066005.
\bibitem{21}
Oikonomou, V. K. "Rescaled Einstein-Hilbert gravity from f (R) gravity: Inflation, dark energy, and the swampland criteria." Physical Review D 103.12 (2021): 124028.
\bibitem{22}
Sadeghi, J., et al. "de Sitter Swampland Conjecture in String Field Inflation." The European Physical Journal C 83 (2023) : (635).
\bibitem{23}
Harlow, Daniel, et al. "Weak gravity conjecture." Reviews of Modern Physics 95.3 (2023): 035003.
\bibitem{24}
Capozziello, Salvatore, et al. "Hydrostatic equilibrium and stellar structure in f (R) gravity." Physical Review D 83.6 (2011): 064004.
\bibitem{25}
Gashti, S. Noori, et al. "Swampland dS conjecture in mimetic f (R, T) gravity." Communications in Theoretical Physics 74.8 (2022): 085402.
\bibitem{26}
Das, Suratna. "Distance, de Sitter and Trans-Planckian Censorship conjectures: the status quo of Warm Inflation." Physics of the Dark Universe 27 (2020): 100432.
\bibitem{27}
Yuennan, Jureeporn, and Phongpichit Channuie. "Further Refining Swampland Conjecture on Inflation in General Scalar‐Tensor Theories of Gravity." Fortschritte der Physik 70.6 (2022): 2200024.
\bibitem{28}
Bedroya, Alek, and Cumrun Vafa. "Trans-Planckian censorship and the swampland." Journal of High Energy Physics 2020.9 (2020): 1-34.
\bibitem{29}
Sadeghi, Jafar, et al. "Can black holes cause cosmic expansion?." arXiv preprint arXiv:2305.12545 (2023).
\bibitem{30}
Mohammadi, Abolhassan, Tayeb Golanbari, and Jamil Enayati. "Brane inflation and Trans-Planckian censorship conjecture." Physical Review D 104.12 (2021): 123515.
\bibitem{31}
Sadeghi, Jafar, Mohammad Reza Alipour, and Saeed Noori Gashti. "Scalar Weak Gravity Conjecture in Super Yang-Mills Inflationary Model." Universe 8.12 (2022): 621.
\bibitem{32}
Kallosh, Renata, et al. "dS Vacua and the Swampland." Journal of High Energy Physics 2019.3 (2019): 1-18.
\bibitem{33}
Guleryuz, Omer. "On the Trans-Planckian Censorship Conjecture and the generalized non-minimal coupling." Journal of Cosmology and Astroparticle Physics 2021.11 (2021): 043.
\bibitem{34}
Osses, Constanza, Nelson Videla, and Grigoris Panotopoulos. "Reheating in small-field inflation on the brane: the swampland criteria and observational constraints in light of the PLANCK 2018 results." The European Physical Journal C 81 (2021): 1-29.
\bibitem{35}
Sadeghi, J., S. Noori Gashti, and M. R. Alipour. "Notes on further refining de Sitter swampland conjecture with inflationary models." Chinese Journal of Physics 79 (2022): 490-502.
\bibitem{36}
Brahma, Suddhasattwa. "Trans-Planckian censorship, inflation, and excited initial states for perturbations." Physical Review D 101.2 (2020): 023526.
\bibitem{37}
Brandenberger, Robert. "Trans-Planckian censorship conjecture and early universe cosmology." arXiv preprint arXiv:2102.09641 (2021).
\bibitem{38}
Sadeghi, J., S. Noori Gashti, and F. Darabi. "Swampland conjectures in hybrid metric-Palatini gravity." Physics of the Dark Universe (2022): 101090.
\bibitem{39}
Geng, Hao, Sebastian Grieninger, and Andreas Karch. "Entropy, entanglement and swampland bounds in DS/dS." Journal of High Energy Physics 2019.6 (2019): 1-16.
\bibitem{40}
Gashti, S. Noori, J. Sadeghi, and B. Pourhassan. "Pleasant behavior of swampland conjectures in the face of specific inflationary models." Astroparticle Physics 139 (2022): 102703.
\bibitem{41}
Sadeghi, J., E. Naghd Mezerji, and S. Noori Gashti. "Study of some cosmological parameters in logarithmic corrected f (R) gravitational model with swampland conjectures." Modern Physics Letters A 36.05 (2021): 2150027.
\bibitem{42}
Sadeghi, J., S. Noori Gashti, and E. Naghd Mezerji. "The investigation of universal relation between corrections to entropy and extremality bounds with verification WGC." Physics of the Dark Universe 30 (2020): 100626.
\bibitem{43}
Agrawal, Prateek, et al. "On the cosmological implications of the string swampland." Physics Letters B 784 (2018): 271-276.
\bibitem{44}
Odintsov, Sergei D., and Vasilis K. Oikonomou. "Swampland implications of GW170817-compatible Einstein-Gauss-Bonnet gravity." Physics Letters B 805 (2020): 135437.
\bibitem{45}
Sadeghi, J., and S. Noori Gashti. "Investigating the logarithmic form of f (R) gravity model from brane perspective and swampland criteria." Pramana 95 (2021): 1-8.
\bibitem{46}
Sharma, Umesh Kumar. "Reconstruction of quintessence field for the THDE with swampland correspondence in f (R, T) gravity." International Journal of Geometric Methods in Modern Physics 18.02 (2021): 2150031.
\bibitem{47}
Sadeghi, Jafar, Mohammad Reza Alipour, and Saeed Noori Gashti. "Strong cosmic censorship in light of weak gravity conjecture for charged black holes." Journal of High Energy Physics 2023.2 (2023): 1-14.
\bibitem{48}
Odintsov, S. D., V. K. Oikonomou, and L. Sebastiani. "Unification of constant-roll inflation and dark energy with logarithmic R2-corrected and exponential F (R) gravity." Nuclear Physics B 923 (2017): 608-632.
\bibitem{49}
Sadeghi, J., et al. "de Sitter Swampland Conjecture in String Field Inflation."  The European Physical Journal C 83 (2023): (635).
\bibitem{50}
Shokri, Mehdi, Jafar Sadeghi, and Saeed Noori Gashti. "Quintessential constant-roll inflation." Physics of the Dark Universe 35 (2022): 100923.
\bibitem{51}
Sadeghi, J., et al. "Swampland conjecture and inflation model from brane perspective." Physica Scripta 96.12 (2021): 125317.
\bibitem{52}
Noori Gashti, Saeed, et al. "Exploring the Parameter Space of Inflation Model on the Brane and its Compatibility with the Swampland Conjectures" arXiv preprint arXiv:2409.06488 (2024).10.22128/jhap.2024.852.1089
\bibitem{53}
Sadeghi, J., et al. "The emergence of universal relations in the AdS black holes thermodynamics." Physica Scripta 98.2 (2023): 025305.
\bibitem{54}
Yuennan, Jureeporn, and Phongpichit Channuie. "Composite Inflation and further refining dS swampland conjecture." Nuclear Physics B 986 (2023): 116033.
\bibitem{55}
Gashti, S. Noori, and J. Sadeghi. "Refined swampland conjecture in warm vector hybrid inflationary scenario." The European Physical Journal Plus 137.6 (2022): 1-13.
\bibitem{56}
Kinney, William H. "Eternal inflation and the refined swampland conjecture." Physical review letters 122.8 (2019): 081302.
\bibitem{57}
Kinney, William H. "The swampland conjecture bound conjecture." arXiv preprint arXiv:2103.16583 (2021).
\bibitem{58}
Yu, Ten-Yeh, and Wen-Yu Wen. "Cosmic censorship and weak gravity conjecture in the Einstein–Maxwell-dilaton theory." Physics Letters B 781 (2018): 713-718.
\bibitem{59}
Gashti, S. Noori. "Two-field inflationary model and swampland de Sitter conjecture." Journal of Holography Applications in Physics 2 (1) (2022): 13-24.
\bibitem{60}
Vafa, Cumrun. "The String landscape and the swampland." arXiv preprint hep-th/0509212 (2005).
\bibitem{61}
Sadeghi, Jafar, et al. "Weak gravity conjecture from conformal field theory: a challenge from hyperscaling violating and Kerr-Newman-AdS black holes." Chinese Physics C 47.1 (2023): 015103.
\bibitem{62}
van Beest, Marieke, et al. "Lectures on the swampland program in string compactifications." Physics Reports 989 (2022): 1-50.
\bibitem{63}
Sadeghi, J., et al. "Weak gravity conjecture, black branes and violations of universal thermodynamics relation." Annals of Physics 447 (2022): 169168.
\bibitem{64}
Kolb, Edward W., Andrew J. Long, and Evan McDonough. "Gravitino swampland conjecture." Physical Review Letters 127.13 (2021): 131603.
\bibitem{65}
Yuennan, Jureeporn, and Phongpichit Channuie. "Further Refining Swampland Conjecture on Inflation in General Scalar‐Tensor Theories of Gravity." Fortschritte der Physik 70.6 (2022): 2200024.
\bibitem{66}
Palti, Eran. "The swampland: introduction and review." Fortschritte der Physik 67.6 (2019): 1900037.
\bibitem{67}
Noori Gashti, S., J. Sadeghi, and M. R. Alipour. "Further refining swampland dS conjecture in mimetic f (G) gravity." International Journal of Modern Physics D 32.03 (2023): 2350011.
\bibitem{68}
Alipour, Mohammad Reza, Jafar Sadeghi, and Mehdi Shokri. "WGC and WCC for charged black holes with quintessence and cloud of strings." The European Physical Journal C 83.7 (2023): 1-7.
\bibitem{69}
Sadeghi, Jafar, et al. "Cosmic evolution of the logarithmic f (R) model and the dS swampland conjecture." Universe 8.12 (2022): 623.
\bibitem{70}
Cheung, Clifford, and Grant N. Remmen. "Naturalness and the weak gravity conjecture." Physical review letters 113.5 (2014): 051601.
\bibitem{71}
Anand, Ankit. "Thermodynamic Extremality in Power-law AdS Black Holes A Universal Perspective." arXiv preprint arXiv:2409.07079 (2024).
\bibitem{72}
Sadeghi, Jafar, and Saeed Noori Gashti. "Influences of perfect fluid dark matter on coinciding validity of the weak gravity and weak cosmic censorship conjectures for Kerr-Newman black hole." Nuclear Physics B 1006 (2024): 116657.
\bibitem{73}
Sadeghi, Jafar, and Saeed Noori Gashti. "Reissner-Nordström black holes surrounded by perfect fluid dark matter: Testing the viability of weak gravity conjecture and weak cosmic censorship conjecture simultaneously." Physics Letters B 853 (2024): 138651.
\bibitem{73'}
Alipour, Mohammad Reza, Jafar Sadeghi, and Mehdi Shokri. "WGC and WCCC of black holes with quintessence and cloud strings in RPS space." Nuclear Physics B 990 (2023): 116184.
\bibitem{82}
Alipour, Mohammad Reza, and Jafar Sadeghi. "The interplay of WGC and WCCC via charged scalar field fluxes in the RPST framework." arXiv preprint arXiv:2406.13784 (2024).
\bibitem{74}
Sadeghi, Jafar, et al. "Weak Cosmic Censorship and Weak Gravity Conjectures in CFT Thermodynamics." arXiv preprint arXiv:2404.15981 (2024).
\bibitem{75}
Cong, Wan, et al. "Holographic CFT phase transitions and criticality for charged AdS black holes." Journal of High Energy Physics 2022.8 (2022): 1-37.
\bibitem{76}
Gong, Ting-Feng, Jie Jiang, and Ming Zhang. "Holographic thermodynamics of rotating black holes." Journal of High Energy Physics 2023.6 (2023): 1-22.
\bibitem{77}
Baruah, Abhishek, and Prabwal Phukon. "Holographic CFT thermodynamics of charged, rotating black holes in $ D= 4$ dimension." arXiv preprint arXiv:2407.02997 (2024).
\bibitem{78}
Ahmed, Moaathe Belhaj, et al. "Holographic dual of extended black hole thermodynamics." Physical Review Letters 130.18 (2023): 181401.
\bibitem{79}
Baruah, Abhishek, and Prabwal Phukon. "Holographic CFT phase transitions for 4-D Dyonic AdS Black Holes." arXiv preprint arXiv:2407.11058 (2024).
\bibitem{80}
Ahmed, Moaathe Belhaj, et al. "Holographic CFT phase transitions and criticality for rotating AdS black holes." Journal of High Energy Physics 2023.8 (2023): 1-32.
\bibitem{81}
Ladghami, Yahya, and Taoufik Ouali. "Black holes thermodynamics with CFT re-scaling." Physics of the Dark Universe 44 (2024): 101471.
\bibitem{83}
Shaymatov, Sanjar, Bobomurat Ahmedov, and Mubasher Jamil. "Testing the weak cosmic censorship conjecture for a Reissner–Nordström–de Sitter black hole surrounded by perfect fluid dark matter." The European Physical Journal C 81 (2021): 1-11.
\bibitem{84}
Kong, Lingyao, Daniele Malafarina, and Cosimo Bambi. "Can we observationally test the weak cosmic censorship conjecture?." The European Physical Journal C 74 (2014): 1-12.
\bibitem{85}
Hod, Shahar. "Weak cosmic censorship: as strong as ever." Physical review letters 100.12 (2008): 121101.
\bibitem{86}
Wang, Xin-Yang, and Jie Jiang. "Examining the weak cosmic censorship conjecture of RN-AdS black holes via the new version of the gedanken experiment." Journal of Cosmology and Astroparticle Physics 2020.07 (2020): 052.
\bibitem{87}
Zhao, Min, Meirong Tang, and Zhaoyi Xu. "Testing the weak cosmic censorship conjecture in short haired black holes." The European Physical Journal C 84.5 (2024): 497.
\bibitem{88}
Gwak, Bogeun. ”Weak cosmic censorship with pressure and volume in charged anti-de Sitter black hole
under charged scalar field.” Journal of Cosmology and Astroparticle Physics 2019.08 (2019): 016.
\bibitem{89}
Gwak, Bogeun. ”Weak cosmic censorship conjecture in Kerr-(anti-) de Sitter black hole with scalar field.”
Journal of High Energy Physics 2018.9 (2018): 1-22.
\bibitem{90}
Barausse, Enrico, Vitor Cardoso, and Gaurav Khanna. ”Test bodies and naked singularities: Is the
self-force the cosmic censor?.” Physical Review Letters 105.26 (2010): 261102.
\bibitem{91}
Hod, Shahar. ”Weak cosmic censorship: as strong as ever.” Physical review letters 100.12 (2008): 121101.
\bibitem{92}
Horowitz, Gary T., Jorge E. Santos, and Benson Way. "Evidence for an electrifying violation of cosmic censorship." Classical and Quantum Gravity 33.19 (2016): 195007.
\bibitem{93}
Crisford, Toby, and Jorge E. Santos. "Violating the weak cosmic censorship conjecture in four-dimensional Anti–de Sitter space." Physical Review Letters 118.18 (2017): 181101.
\bibitem{94}
Kiselev, VV1966820. "Quintessence and black holes." Classical and Quantum Gravity 20.6 (2003): 1187.
\bibitem{95}
Cao, Yihe, et al. "Joule–Thomson expansion of RN-AdS black hole immersed in perfect fluid dark matter." Communications in Theoretical Physics 73.9 (2021): 095403.
\bibitem{96}
Li, Ming-Hsun, and Kwei-Chou Yang. "Galactic dark matter in the phantom field." Physical Review D 86.12 (2012): 123015.
\bibitem{97}
Kiselev, V. V. "Quintessential solution of dark matter rotation curves and its simulation by extra dimensions." arXiv preprint gr-qc/0303031 (2003).
\bibitem{98}
Xu, Zhaoyi, Xian Hou, and Jiancheng Wang. "Kerr–anti-de Sitter/de Sitter black hole in perfect fluid dark matter background." Classical and Quantum Gravity 35.11 (2018): 115003.

\bibitem{Bekenstein:1973mi}
J.~D.~Bekenstein, ``Extraction of energy and charge from a black hole,'' Phys. Rev. D \textbf{7} (1973), 949-953 doi:10.1103/PhysRevD.7.949.

\bibitem{Brito:2015oca}
R.~Brito, V.~Cardoso and P.~Pani, "Superradiance: Energy Extraction, Black-Hole Bombs and Implications for Astrophysics and Particle Physics", Lect.\ Notes Phys.\  {\bf 906}, pp.1 (2015), [arXiv:1501.06570 [gr-qc]]
\end{thebibliography}
\end{document}